\documentclass[aps,prl,a4paper,tightenlines,amsmath,amssymb,superscriptaddress,twocolumn,floatfix,final]{revtex4}
\usepackage{latexsym} \usepackage{amssymb}
\usepackage[dvips]{graphics} \usepackage[pdftex]{graphicx}

 \newcommand{\beq}{\begin{eqnarray}}
\newcommand{\eeq}{\end{eqnarray}} \newcommand{\be}{\begin{eqnarray}}
\newcommand{\ee}{\end{eqnarray}}

\usepackage[latin1]{inputenc} \usepackage{amsmath}
\usepackage{amsfonts} \usepackage{amssymb} \usepackage{amsthm}
\begin{document}
\title{Strongly Coupled Chameleon Fields: New Horizons in Scalar Field Theory}
\author{David F. Mota}
\affiliation{Institute of Theoretical Astrophysics, University of Oslo
N-0315, Oslo, Norway}
\affiliation{Perimeter Institute for Theoretical
Physics, Waterloo, Ontario N2L 2Y5, Canada}
\author{Douglas J. Shaw}
\affiliation{DAMTP, Centre for Mathematical Sciences, University of
Cambridge, Wilberforce Road, Cambridge CB3 0WA, UK}
\begin{abstract}
We show that as a result of non-linear self-interactions, scalar field
theories that couple to matter much more strongly than gravity are not
only viable but could well be detected by a number of future
experiments provided they are properly designed to do so. 
%{\bf{shall we add something else?}}
\end{abstract}
\maketitle
There is wide-spread interest in the possibility that, in addition
to the matter described by the standard model of particle physics,
our Universe may be populated by one or more scalar fields.
%{\footnote{in addition to the Higgs scalar of the standardmodel}} 
These are a general feature in high energy physics beyond
the standard model and are often related to the presence of
extra-dimensions. The existence of scalar fields has also been
postulated as means to explain the early and late time
acceleration of the Universe.
%As well as any possible variation in the
%fundamental constants of nature.
%
%When scalar fields do exist,
It is almost always the case that such fields interact with
matter: either due to a direct Lagrangian coupling or indirectly
through a coupling to the Ricci scalar or as the result of quantum
loop corrections.  If the scalar field self-interactions are
negligible, then the experimental bounds on such a field are very
strong: requiring it to either couple to matter much more weakly
than gravity does, or to be very heavy \cite{bounds}.
%In either case, the scalar field
%would have had little or no detectable effect on the history of
%Universe, and would generally be indistinguishable from a cosmological
%constant
%
Recently, a novel scenario was presented by Khoury and Weltman
\cite{cham} that employed self-interactions of the scalar-field to
avoid the most restrictive of the current bounds. They dubbed such
scalars to be `chameleon fields' due to the way in which the
field's mass depends on the density of matter in the local
environment. A chameleon field might be very heavy in relatively
high density environments, such as the Earth and its atmosphere,
but almost massless cosmologically where the density is some
$10^{-30}$ times lower. This feature allows the field to evade
local constraints on fifth force effects and deemed the
\emph{chameleon mechanism}.

Chameleon field theories involve non-linear self-interactions,
%potentials for the field, 
which makes finding analytical solutions 
%of the field equations 
difficult, particularly in highly inhomogeneous
environments. Most commentators invariably, therefore, linearise
the chameleon theories when studying their behaviour in such
backgrounds \cite{cham,othercham}.
%Notably, this was the approach taken in \cite{cham}.
%
In this Letter,
%\cite{motashaw}
%we undertake an in-depth study of generalised
%chameleon theories, and
we show that this linearisation procedure is often invalid. When
properly accounted for, the non-linearities increase the strength of the chameleon mechanism: further hiding
the field from present day constraints, particularly those on
possible violations of the Weak Equivalence Principle (WEP). Our
results not only reveal interesting behaviour at the level of
field theory, but that today's experimental bounds on the
parameters of these theories could be much weaker than previously
realised. Furthermore, they imply that experiments which probe
possible violations of the WEP should be
redesigned if they are to have chance of detecting chameleon
fields.

We consider theories where the chameleon field, $\phi$, has a
self-interaction potential given by: $$ V(\phi) = \lambda M^4
\left(M/\phi\right)^{n}, $$ where $M$ has units of mass,
$n$ is some integer and $\lambda$ is a parameter. We set $c= \hbar=1$ and
define $G= M_{pl}^{-2}$. Theories with $n
>0$ were first consider in this context in \cite{cham}, whilst
$\phi^4$ theory was initially noted to have chameleon-like
behaviour in \cite{phi4}. When $n\neq-4$, we can, by re-scaling
$M$, set $\lambda=1$, whereas when $n = -4$ 
%(i.e. $V \sim \phi^4$)
the mass-scale $M$ does not appear in $V$. As argued in \cite{phi4},
$\lambda=1/4!$ would be a `natural' value when $n=-4$. If $M \sim (0.1\, \mathrm{mm})^{-1}$ the chameleon may play the role of dark energy \cite{othercham}.
% It is often assumed that $M\sim
%1\,\mathrm{eV} \approx 1\, \mathrm{mm}^{-1}$ so that $M^4$ is the
%same order of magnitude as the cosmological vacuum energy density
%today, \cite{cham,othercham}. 

We parameterise the matter coupling of the chameleon by a function
$\beta B_{,\phi}(\beta \phi/M_{pl}) \rho /M_{pl}$. 
%where $M_{pl}$ is the Planck mass. We note below, and in
%\cite{motashaw}, that 
Astrophysical constraints require that
$\vert \beta \phi / M_{pl} \vert \lesssim 0.1$ since
nucleosynthesis \cite{othercham}. Preempting this requirement we simplify our
calculations by expanding $B_{,\phi}$ about $\phi=0$, and scale
$\beta$ so that $B_{,\phi}(0)=1$.  The equation of motion for $\phi$ then becomes
%With this simplification, $\phi$
%obeys the conservation equation:
\begin{eqnarray}
-\square \phi = V_{,\phi}(\phi) + \beta(\rho + \omega
 P)/M_{pl}, \label{field}
\end{eqnarray}
where $\rho$ is the energy density of matter, $P$ is its pressure and
$\omega$ parameterises the way in which the chameleon couples to
matter.  In the simplest models, $\phi$ couples to the trace of the
energy momentum tensor, and so $\omega = -3$. In
what follows we take this to be our fiducial value of $\omega$ and
note that the results for different $O(1)$ values of $\omega$ are very
similar \cite{motashaw}. We note that the right hand side of eq. (\ref{field}) vanishes
when $\phi = \phi_c(\rho +\omega P)$: $$ \phi_c(\rho+\omega P) =
M\left(\beta(\rho + \omega
P)/(\lambda n M_{pl}M^3)\right)^{-\frac{1}{n+1}}. $$ For $\phi_c(\rho + \omega P)$ to be
real when $\beta(\rho + \omega P) > 0$, we need either $n \geq 0$
or for $n$ to be negative and even; and  $n\neq 0, -2$ for the theory to be non-linear. The mass of small perturbations about $\phi = \phi_c$ is $m_c = \sqrt{V_{,\phi \phi}(\phi_c)} = \sqrt{\lambda
n(n+1)}M\left\vert M/\phi_c\right\vert^{n/2 +1}$.

% With these definitions 
One would expect, in the absence of any
chameleon mechanism, the force mediated by $\phi$ to be $\beta^2$
as strong as gravity.  As a result of quantum corrections $\beta$
will generally differ slightly for different particle species, which would standardly lead to a composition dependent
force that would in turn violate WEP. Solar system bounds on WEP
violation require $\beta \lesssim 10^{-5}$ in non-chameleon
theories \cite{bounds}. Chameleon theories have been shown to be
compatible with $\beta \sim O(1)$ \cite{cham}.  
In this Letter,
however, we will go much further and report how, as a result of
non-linear effects, it is possible for a chameleon field to couple
to matter much more strongly than gravity does (i.e $\beta \gg 1$)
and yet for it to have remained thus undetected. We define $M_{\phi} =M_{pl}/\beta$, which is
roughly the energy at which chameleon particles would be produced
in particle colliders.  It would be pleasant in the light of the
hierarchy problem if $M_{pl}/\beta \ll M_{pl}$, say of the GUT
scale, or, if we hoped to find traces of it at the LHC, maybe even
the TeV-scale. We show below that both of these scenarios are
allowed for.

Crucial to our ability to
%to accurately 
constrain chameleon theories
is a full understanding of how they behave as field theories. It
transpires that when $\beta \gg 1$, the non-linear nature of the
potential, $V(\phi)$, becomes very important. Even in the, supposedly,
simple case of the field produced by a single large body, there
might not exist any self-consistent linearisation of eq.
(\ref{field}) that is valid everywhere \cite{motashaw}. Non-linear
effects are also non-negligible when calculating the force
produced by one body upon another. When linearised theory fails, the solution to the two body problem cannot be found simply by superimposing two
copies of the field produced a single body. 

Non-linear effects also play a r\^{o}le
in determining the effective, large-scale or macroscopic theory
associated with the chameleon. Eq. (\ref{field}) defines the
microscopic, or particle-level, field theory for $\phi$, whereas
in most cases we are interested in the large scale or coarse
grained behaviour of $\phi$. In macroscopic bodies the density is
actually strongly peaked near the nuclei of the individual atoms
from which it is formed and these are separated from each other by
distances much greater than their radii. Rather than explicitly
considering the microscopic structure of a body, it is standard
practice to define an `averaged' field theory that is valid over
scales comparable to the body's size. If our field theory were
linear then the averaged equations would be the same as the
microscopic ones e.g. as in Newtonian gravity. But it is important
to note that this is very much a property of \emph{linear
theories} and is not in general true of non-linear ones.
Non-linear effects must, therefore, be taken into account. We do
this by combining matched asymptotic expansions
with exact analytical solution of the full non-linear equations
under certain reasonable assumptions. We confirm our results by
numerically integrating the field equations.
%We find some surprising results.

Firstly we define the concept of a \emph{thin-shell}.  A body is
said to have a thin-shell if the coarse-grained value of $\phi$
(as defined on scales that are large compared to the
sizes of the constituent particles of the body) is approximately
constant everywhere inside the body, except in a thin-shell near
the surface of the body where large changes ($O(1)$) in its value
occur.  The existence of a thin-shell is related to the presence
of non-linear behaviour. Deep inside a body with a thin- shell
$\phi$ is constant, and so we might expect $\phi = \phi_c(\rho)$,
where $\rho$ is the density of the body (we assume $P \ll \rho$).
The effective chameleon mass, $m_{eff}$, in the body would then be
given by $m_{eff}=m_c(\rho)$. The effect of the non-linearities on
the averaging, however, is to limit the averaged value of
$m_{\phi}$ to be smaller than some critical value, $m_{crit}$
\cite{motashaw}. $m_{crit}$ is a macroscopic quantity but it
depends only on the microscopic properties of the body and the
index $n$.  It is \emph{independent} of $\beta$, $M$ and
$\lambda$ \cite{motashaw}.  We have modeled the body as being composed of
particles of radius $R_{p}$ separated by an average distance
$d_{p}$.  The macroscopic mass of the chameleon in the body is
then $m_{eff} = \min\left(m_{c}(\rho), m_{crit}\right)$, where: $$
m_{crit} \approx \sqrt{3\vert n+
1\vert} d_{p}^{-1}\left(R_{p}/d_{p}\right)^{\frac{q(n)}{2}},
\, n \neq -4, $$ where $q(n) = \min(1, (n+4)/(n+1))$ and
$m_{crit} \approx 1.4/d_p$ when $n=-4$.  Whenever $m_{eff}=m_{crit}$ is it because the individual particles that make-up the body have themselves developed thin-shells. This critical behaviour emerges from the requirement that non-linear effects are negligble outside of the particle from $r=R_{p}$ to $r=d_{p}$: this implies a maximal value of $m_{eff}$, i.e. $m_{crit}$, that depends only on $R_{p}$, $d_{p}$ and $n$. The $n$ dependance arises because $n$ determines precisely when linear theory breaksdown.

$\beta$-independent critical behaviour is also seen in the $\phi$-force
between two bodies. The onset of this critical behaviour is linked to the
emergence of a thin-shell. A body of radius $R$ and density $\rho_c$ in a
background of density $\rho_{b} \ll \rho_{c}$ has a thin-shell if:
\be m_{eff}R \gtrsim \sqrt{3\vert n+1 \vert}\left\vert 1-
\left(\rho_c/\rho_b\right)^{\frac{1}{n+1}}\right\vert^{1/2},
n \neq -4. \label{thinshell} \ee The existence of a thin-shell is
essentially due to non-linearities being strong near the surface
of a body but weak in other regions.  % When $m_{eff}R$ is greater
% than $\sqrt{2/3}$ times the right hand side of above expression
% non-linearities begin to play a r\^{o}le near the centre of the
%body.
When $n=-4$, a thin-shell occurs for $m_{eff}R \gtrsim 4$, whereas
linearised theory fails to be accurate for $m_{eff}R \gtrsim 1.4$.
%$m_{eff} = \min(m_c(\rho_c), m_{crit})$ as above.
When $n > 0$, $(\rho_c/\rho_b)^{1/n+1} \gg 1$ and so the
thin-shell condition, eq.(\ref{thinshell}), depends greatly upon
on the background density. The same is \emph{not} true when $n
\leq -4$ since here $(\rho_c/\rho_b)^{(1/n+1)} \ll 1$.
Therefore $n > 0$ theories can behave differently in space-based
experiments than they do in laboratory ones, because the
thin-shell condition is more restrictive in low-density background
of space than it is in the lab\cite{cham}. 
%- a fact first noted in
In contrast, theories with $n \leq -4$ will exhibit
no big difference in their behaviour in space-based tests to
that seen on Earth.

The existence of a thin-shell in the test-masses used in experimental searches for deviations from general relativity is vital if we are to evade their bounds.
Whereas the force between two non-thin-shelled bodies with separation $r$ is $\beta^2 (1+m_b r)e^{-m_b r}$ times the
gravitational force between them ($m_b$ is the chameleon mass in the region between the bodies), the force between two
bodies, of masses $M_1$ and $M_2$, with thin-shells is found to be
independent of the coupling $\beta$ \cite{motashaw}.  When $d \gg R_1, R_2$, where
$R_1$ and $R_2$ are the respective radii of the two bodies, this
force is found to be $\alpha_{12}$ times the strength of gravity,
where for $n \neq -4$: $$ \alpha_{12} = \frac{S(n,m_b)
M_{pl}^2(1+m_b r)e^{-m_b r}}{M_1 M_2} (M^2R_1R_2)^{q(n)}, $$ where
%$q(n)$ is as defined above and
$S(n,m_b)$ is $(3/\vert n \vert)^{2/\vert n + 2 \vert}$ for $n <
-4$, whereas for $n > 0$ it equals $(n(n+1)M^2/m_b^2)^{2/(n+2)}$.
When $n=-4$: $$ \alpha_{12} = \frac{M_{pl}^2(1+m_b r)e^{-m_b
r}}{8\lambda M_1 M_2\sqrt{\ln(r/R_1)\ln(r/R_2)}}. $$ For
$d \lesssim R_1, R_2$ a different value for $\alpha_{12}$ applies
and is given below. This $\beta$
independence was first noted in \cite{nelson}, in the context of $\phi^4$ theory.
However, the authors were mostly
concerned with region of parameter space $\beta < 1$, $\lambda \ll
1$; in our analysis we go further: considering a wider range of
theories and also the possibility that $\beta \gg 1$.

We can understand the $\beta$-independence as follows: just outside a
thin-shelled body, the potential term in eq. (\ref{field}) is large
and negative ($\sim O(-\beta \rho / M_{pl})$), and it causes $\phi$ to
decay very quickly. At some point $\phi$ will reach a critical value,
$\phi_{crit}$, that is small enough so that non-linearities are no
longer important.  Since this all occurs outside the body,
$\phi_{crit}$ can only depend on the size of the body, the choice of
potential $(M, \lambda, n)$ and the mass of $\phi$ in the background,
$m_{b}$. This is precisely what was found above.

This $\beta$-independence is of great of importance if one wishes to
design an experiment to detect the chameleon through WEP
violations. Since the $\phi$-force is independent of the coupling,
$\beta$, for bodies with thin-shells, any microscopic composition
dependence in $\beta$ will be hidden on macroscopic length scales.
The only `composition' dependence in $\alpha_{12}$ is through the
masses of the bodies and their dimensions ($R_1$ and $R_2$). The
strength of WEP violations is quantified by the E\"{o}tvos parameter,
$\eta$. If we measure the differential accelerations of two test
masses, $M_{1}$ and $M_{2}$, of radii $R_{1}$ and $R_{2}$ towards a
third body, mass $M_{3}$ and radius $R_{3}$, then: $\eta =
\alpha_{13}-\alpha_{23}$.  Taking the third body to be the Sun or the
Moon, experiments searches for WEP violations have up to date found
that $\eta \lesssim 10^{-13}$ \cite{WEP}. In most of these searches,
although the composition of the test-masses is different, they are
made to have the \emph{same} mass ($M_1=M_2$) and the \emph{same} size
($R_1 = R_2$).  Therefore, if the test-masses have thin-shells we have
$\eta = 0$ and \emph{no} WEP violation will be detected.  The only
implicit dependence of this result on $\beta$ is that the
\emph{larger} the coupling is, the more likely it is that the
test-masses will satisfy the thin-shells conditions. The first
important consideration for future experiments is that: if one wishes
to detect a chameleon field through WEP violations one must either
ensure that test-masses do not satisfy the thin-shell conditions or that
they are of different masses and/or dimensions.

\begin{figure}[tbh]
\begin{center}
\includegraphics[scale=0.357]{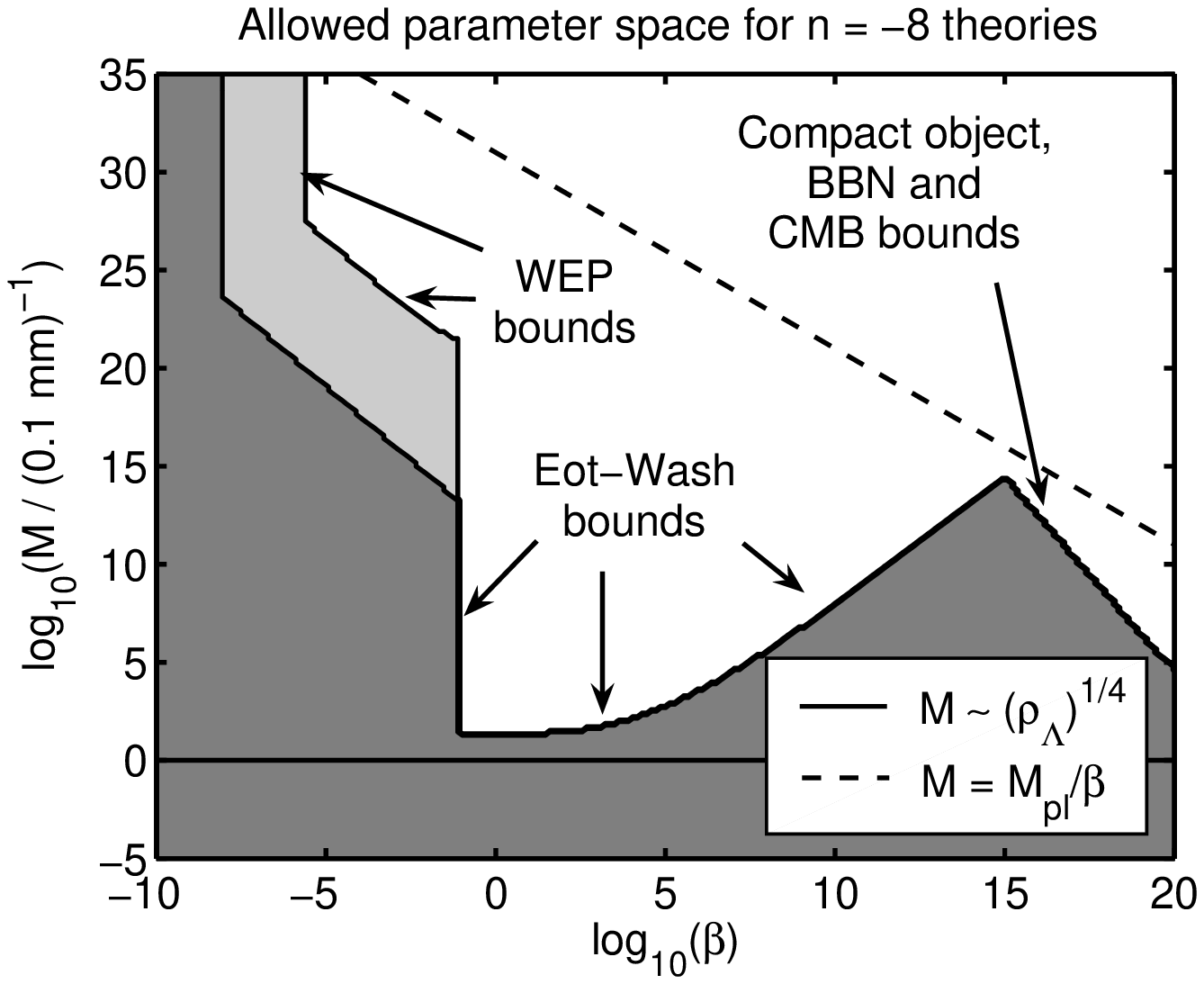} 
\includegraphics[scale=0.357]{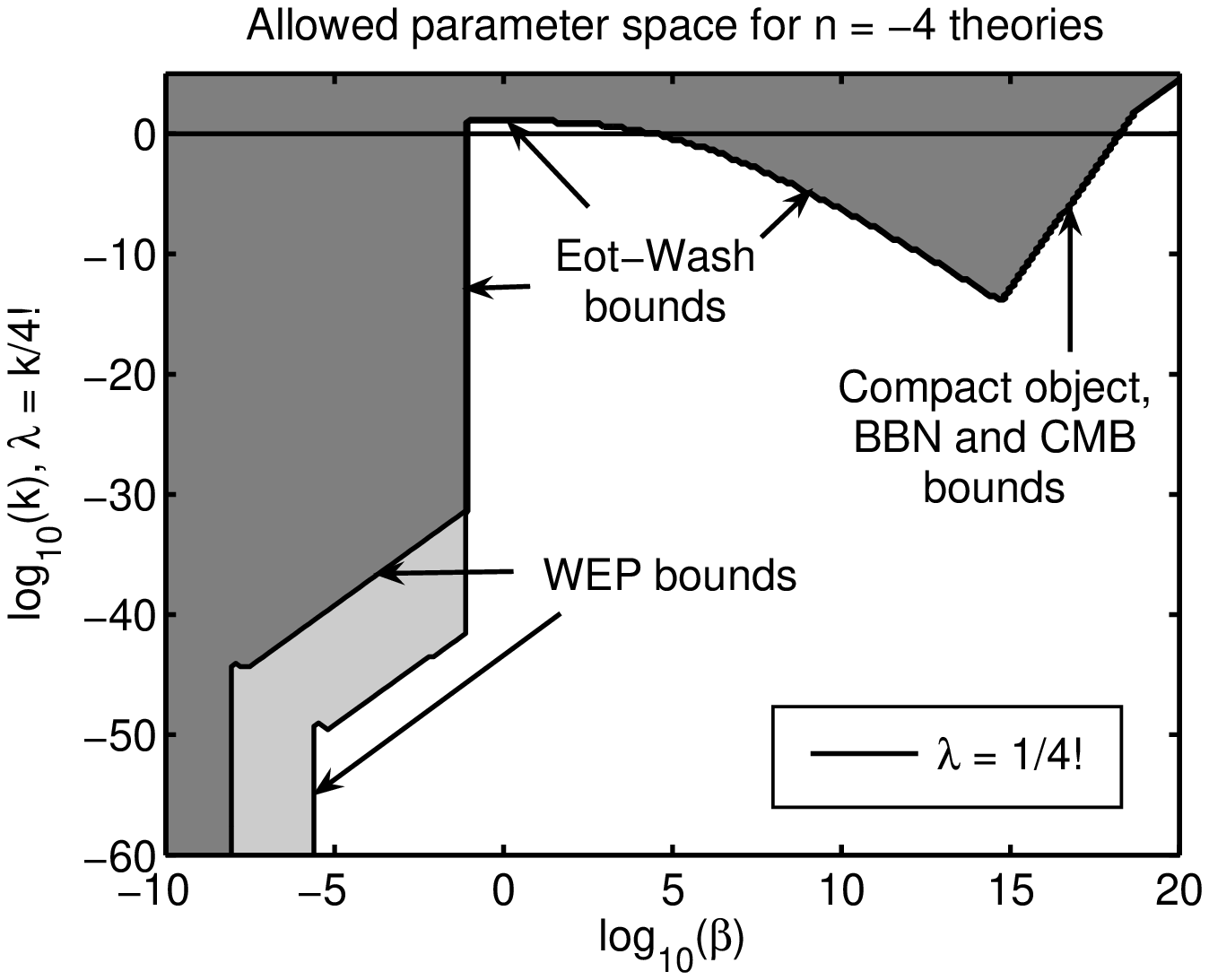}
\includegraphics[scale=0.357]{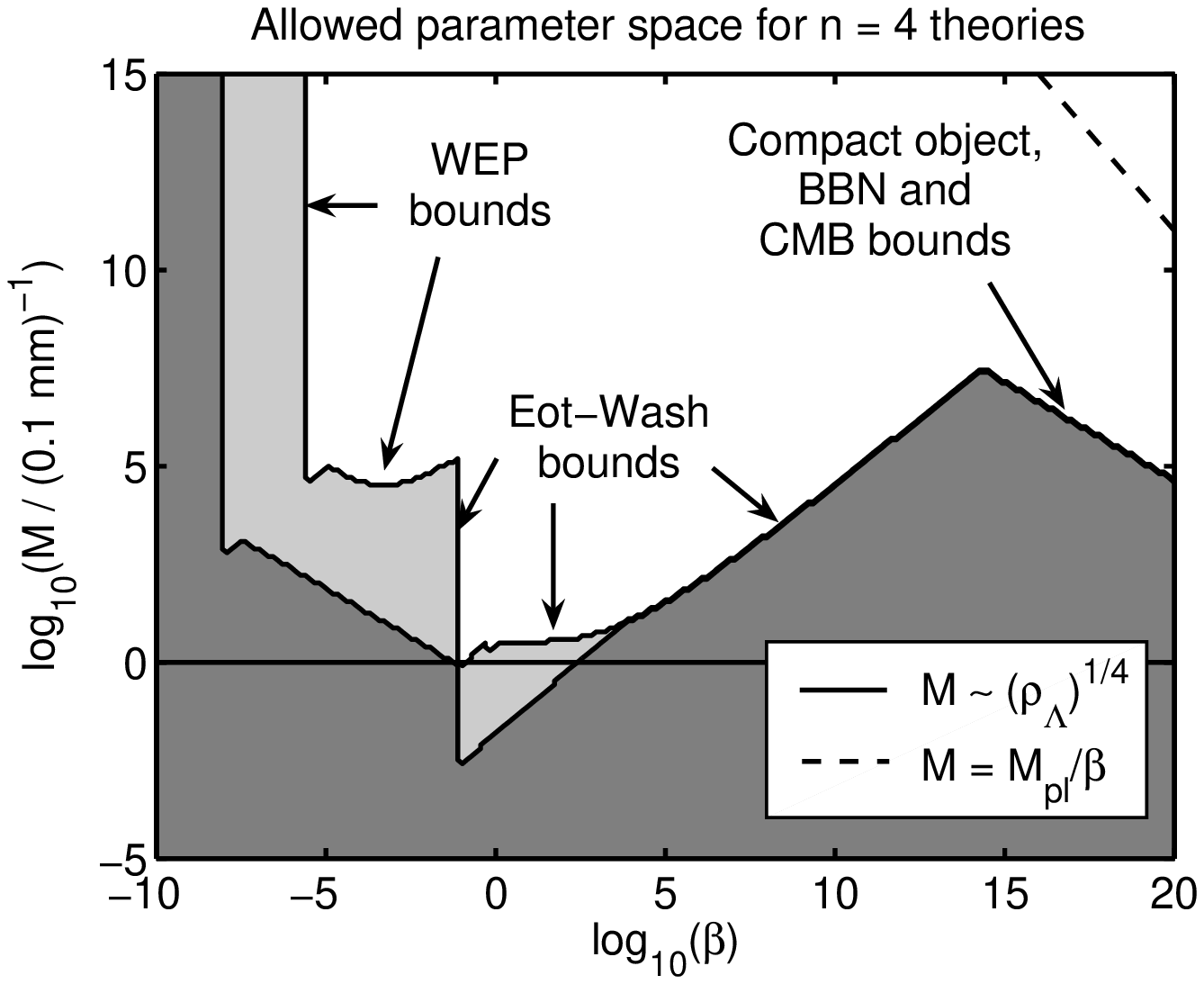}
\end{center}
\caption{The whole of shaded area shows the allowed parameter space
with all current bounds. For some values of $M$ and $\lambda$ we need
the BeCu sheet to have a thin-shell; this results in a region near
$\beta \sim O(1)$ being ruled out. Future space-based tests could
detect the more lightly shaded regions.  The solid horizontal lines
indicates the case where the chameleon field behaves like dark energy.
%We do not plot past $\beta = 10^{-6}$ as this is already known to be
%permissible and past $\beta = 10^{19} \Rightarrow M_{pl}/\beta \sim
%1\,\mathrm{GeV}$ as this is probably ruled out by particle
%colliders. 
Plots for theories with $n < -4$ or $n > 0$ are
similar to cases $n = -8$ and $n = 4$ respectively.}
\label{fig1}
\end{figure}
We shall assume that such an experiment as been conducted, using two
spherical test bodies both with a mass of $10\,g$, where one is made
entirely of copper and the other of aluminum.  The strongest bounds on
chameleon fields would then come from measuring the differential
acceleration of these bodies towards the Moon.  We indicate in
FIG.\ref{fig1} the restrictions that finding $\eta \lesssim 10^{-13}$
in such an experiment would place on these chameleon theories.  The
Moon is a better choice of attractor than the Earth or the Sun for
such experiments since $\alpha_{13}$ is proportional to
$M_{pl}^2/M_{1}M_{3}$ and so the smaller mass of the test-bodies,
$M_{1}$, and the attractor, $M_{3}$, the larger $\eta$ will be
compared to gravity.  The corollary of this result is that if we are
unable to detect $\phi$ in lab-based, micro-gravity experiments where
both $M_1$ and $M_2 \sim O(10 \, g)$ (such as the E\"{o}t-Wash
experiment) then the $\phi$-force between larger (say human-sized)
objects, would also be undetectably small.  For this reason
measurements of the differential acceleration of the Earth and Moon
towards the Sun, e.g. lunar laser ranging, are not competitive with
lab-based experiments.

Future, space-based tests of WEP promise to be able to detect $\eta$
up to a precision of $10^{-18}$; we indicate on FIG.\ref{fig1}, the
regions of parameter space that such experiments would be able to
detect.
The $\phi$-mediated force will also produce effective corrections to
the $1/r^2$ behaviour of gravity.  The best bounds on such corrections
come from the E\"{o}t-Wash experiment performed by Hoyle et
al\cite{EotWash} which employs a torsion balance to measure the torque
induced on a pendulum by a rotating attractor at a separation $d$.
For $d \gtrsim 0.1 \mathrm{mm}$, they find that $\alpha_{12} \lesssim
10^{-2}$ {\cite{EotWash}}. For a chameleon theory to satisfy this
bound we need both the attractor and pendulum to have thin-shells. In
this scenario $d$ is small compared to the size of test-masses ($d <
R_{1}, R_{2}$) and so the previous formula for $\alpha_{12}$ does not
apply. When the mass of the chameleon inside the attractor and
pendulum, $m_{\phi}$, obeys $m_{\phi}d \gg 1$ (as is the case for
$\beta \gtrsim 1$) we find that the $\phi$-force is $\alpha$ times the
strength of gravity, where $\alpha_{12}$ is: $$ 5 \times
10^{-4}\left(\frac{M}{(0.1 \,
\mathrm{mm})^{-1}}\right)^{\frac{2(n+4)}{n+2}}\left(\frac{\lambda^{1/n}\sqrt{2}B\left(\frac{1}{2},\frac{1}{2}+\frac{1}{n}\right)}{\vert
n \vert d /\, 0.1 \, \mathrm{mm}}\right)^{\frac{2n}{n+2}}, $$ where
$B(p,q)$ is the beta function. We note that $\alpha$, as before, is
\emph{independent} of $\beta$. The E\"{o}t-Wash bound is strongest for
$n=-4$ where it appears to rule out a `natural' value for $\lambda$ of
$1/4!$: $0.56 \lambda^{-1} \lesssim 1$.  However this is not the whole
story. In this experiment a uniform $10\mu m$ thick
%{\footnote{now reduced to be $10 \mu m$ thick}}
BeCu membrane is placed between the pendulum and attractor to shield
electromagnetic forces.  For $O(1)$ values of $\beta$ and $\lambda
\sim 1/4!$ or $M \sim (0.1 \mathrm{mm})^{-1}$ this sheet does not have a
thin-shell and makes little difference to the analysis.  For slightly
larger values of $\beta$ however ($\beta \gtrsim 10^{4}$ and $\lambda
= 1/4!$ for $n=-4$) it will develop a thin-shell.  Taking the mass of
the chameleon inside the sheet to be $m_{s}$, the effect of this
membrane is then to attenuate $\alpha_{12}$ by a factor of
$\exp(-m_{s}d_{s})$, where $d_{s}$ is the thickness of the sheet.  The
E\"{o}t-Wash bound is then easily satisfied even for $\lambda \sim
1/4!$.  The larger $\beta$ becomes, the larger $m_{s}$ is and the less
restrictive this bound becomes.  Experiments such as this must
therefore be redesigned if they are to be able to detect chameleon
theories with $\beta \gg 1$.
%In figure \ref{fig1} we plot the allowed parameter space for a theory
%with $n = -8$, $n=-4$, and $n = 4$ {\bf{Is this with BeCu?}}.  

The prospect that couplings with $\beta \gg 1$ could be allowed is
exciting. But to be taken seriously we must also consider bounds
coming from astrophysical constraints, such as the stability and
mass-radius relationship of white dwarfs and neutron stars as well as
bounds coming from big bang nucleosynthesis (BBN) and the Cosmic
Microwave Radiation temperature anisotropies
\cite{motashaw,othercham}.  These bounds can be summarised as
requiring $\vert\beta \phi / M_{pl}\vert \lesssim 0.1$ over the whole
universe since the BBN epoch \cite{motashaw, othercham}. This
condition is enough to ensure that there has been no more than a 10\%
change in particle masses since BBN and in the redshift of the surface
of last-scattering.
%We note that the constraints on $\beta$ found in
%\cite{CMB1} do not apply because $\phi$ is not `light' cosmologically
%as assumed in that work. Also structure formation constraints are not competitive with the other astrophysical bounds.
Whilst we satisfy the same physical constraints as Amendola for
non-chameleon, coupled quintesence \cite{CMB1}, the chameleon
mechanism ensures a significantly less restrictive bound on $\beta$
than was found there.  Astrophysical constraints only place a weak
upper bound on $\beta$ which is strongest for $n=-4$, e.g. if $\lambda
= 1/4!$ we need $M_{pl}/\beta \gtrsim 10\,\mathrm{GeV}$. However,
realistically, we probably require $M_{pl}/\beta \gtrsim 200\,
\mathrm{GeV}$ for it not to have been seen so far in particle
colliders. 
%A full study of the effect of $\phi$ on scattering
%amplitudes would require a quantum theory for the chameleon which
%falls outside of the scope of this work.

In summary, we have considered a wide spectrum of scalar field
theories with a chameleon mechanism and for the first time, the
non-linear structure of these theories has been properly taken into
account.  We have found a surprising result that the chameleon force
between two bodies with thin-shell is \emph{independent} of their
coupling to the field $\phi$, and that as a result the bounds on the
coupling, $\beta$, can be exponentially relaxed. We have also noted
that some laboratory experiments should be redesigned to detect the
chameleon. For `natural' values of $M \sim (0.1 \mathrm{mm})^{-1}$ or
$\lambda \sim 1/4!$, the strongest upper bounds on $\beta$ probably
come from particle colliders and $200 \mathrm{GeV} \lesssim
M_{pl}/\beta \lesssim 10^{15} \mathrm{GeV}$ is allowed for all $n$.
If $M_{pl}/\beta \sim 1 \mathrm{TeV}$ we might even hope to see
chameleon production at the LHC; although without a renormalisable quantum theory of the chameleon it is hard to say for sure if this happen.   Planned space-based tests such as
STEP, MICROSCOPE and SEE, \cite{space}, promise improved precision
and, when $n > 0$ there is also still the possibility that WEP
violations in space can be stronger than the level already ruled out
by laboratory based experiments.  As noted in \cite{cham,othercham},
the chameleon field is a good candidate for dark energy if $M \sim
(\rho_{\Lambda})^{1/4} \approx (0.1\, \mathrm{mm})^{-1}$; this result is
unchanged for $\beta \gg 1$.

In conclusion: scalar field theories that couple to matter much more
strongly than gravity are not only viable but could well be detected
by a number of future experiments provided they are properly designed
to do so.  This result opens up an altogether new window which might
lead to a completely different view of the r\^{o}le played by scalar
fields in particle physics and cosmology.
\acknowledgments DFM acknowledges the Research Council of Norway and
Perimeter Institute. DJS acknowledges PPARC.

\begin{thebibliography}{99}
%
\bibitem{cham} J.~Khoury and A.~Weltman, Phys. Rev. D {\bf 69}, 044026
  (2004); Phys. Rev. Lett. {\bf 93} 171104 (2004).
%
\bibitem{othercham} Ph.~Brax {\it et al.}, Phys. Rev. D {\bf 70}
123518 (2004).

\bibitem{motashaw} D.~F.~Mota and D.~J.~Shaw, hep-ph/0608078.

%
\bibitem{phi4} S.Gubser and J.Khoury, Phys.Rev.D{\bf 70},104001(2004).
%
\bibitem{space} A.~J.~Sanders {\it et al.}, Meas Sci. Technol. {\bf
19}, 514 (199); P.~Touboul {\it et al.}, Acta Astronaut. {\bf 50}, 433
(2002); A.~M.~Nobili {\it et al.}, Class. Quant. Grav. {\bf 17}, 2347
(2000); J.~Mester {\it et al.} Class. Quant. Grav. {\bf 18}, 2475
(2001).
%
\bibitem{WEP} S.~Orito {\it et al.}, Phys. Rev. Lett. {\bf 84}, 1078
(2000); Y.~Asaoka {\it et al.}, Phys. Rev. Lett. {\bf 88}, 51101
(2002).
%
\bibitem{EotWash} C.~D.~Hoyle {\it et al.}, Phys. Rev. Lett. {\bf 86},
1418 (2001).

\bibitem{bounds} J.~P.~Uzan, Rev. Mod. Phys. {\bf 75}, 403 (2003); B.~Bertotti {\it et al.}, Nature {\bf 425}, 374 (2003).

\bibitem{nelson} B.~Feldman and A.~Nelson, hep-ph/0603057.
%
\bibitem{CMB1} L.~Amendola,  Phys. Rev. D {\bf 62}, 043511 (2000).
\end{thebibliography}
\end{document}